\documentclass[final,authoryear,5p,times,twocolumn]{elsarticle}

\usepackage{gensymb}

\usepackage[pdftex,pdfpagemode={UseOutlines},bookmarks,bookmarksopen,colorlinks,linkcolor={blue},citecolor={green},urlcolor={red}]{hyperref}

\usepackage{breakurl}

\usepackage{listings}

\journal{Astronomy \& Computing}

\usepackage{upquote}

\usepackage{color}

\begin{document}

\begin{frontmatter}



\title{Learning from FITS: Limitations in use in modern astronomical research}


\author[noao]{Brian~Thomas\corref{cor1}}
\ead{bthomas@noao.edu}
\author[cornell]{Tim~Jenness}
\author[noao]{Frossie~Economou}
\author[stsci]{Perry~Greenfield}
\author[geminin]{Paul~Hirst}
\author[jac]{David~S.~Berry}
\author[stsci]{Erik~Bray}
\author[glasgow]{Norman~Gray}
\author[ohio]{Demitri~Muna}
\author[geminis]{James~Turner}
\author[princeton]{Miguel~de~Val-Borro}
\author[iaa,ska]{Juande~Santander-Vela}
\author[ipac]{David~Shupe}
\author[ipac]{John~Good}
\author[ipac]{G.~Bruce~Berriman}
\author[icrar]{Slava~Kitaeff}
\author[microsoft]{Jonathan~Fay}
\author[sao]{Omar~Laurino}
\author[stsci]{Anastasia~Alexov}
\author[ipac]{Walter~Landry}
\author[nrao]{Joe~Masters}
\author[cornell]{Adam~Brazier}
\author[aifa]{Reinhold~Schaaf}
\author[uwaterloo]{Kevin~Edwards}
\author[jac]{Russell~O.~Redman}
\author[warwick]{Thomas~R.~Marsh}
\author[aip]{Ole~Streicher}
\author[noao]{Pat~Norris}
\author[ucm]{Sergio~Pascual}
\author[unsw]{Matthew~Davie}
\author[stsci]{Michael~Droettboom}
\author[mpia]{Thomas~Robitaille}
\author[iasf]{Riccardo~Campana}
\author[psu]{Alex~Hagen}
\author[mps]{Paul~Hartogh}
\author[aifa]{Dominik~Klaes}
\author[msum]{Matthew~W.~Craig}
\author[cral]{Derek~Homeier}

\cortext[cor1]{Corresponding author}

\address[noao]{Science Data Management, National Optical Astronomy Observatory, 950 N Cherry Ave, Tucson, AZ 85719, USA}
\address[cornell]{Department of Astronomy, Cornell University, Ithaca,
  NY 14853, USA}
\address[stsci]{Space Telescope Science Institute, 3700 San Martin Drive, Baltimore, MD 21218, USA}
\address[geminin]{Gemini Observatory, 670 N.\ A`oh\=ok\=u Place, Hilo, HI 96720, USA}
\address[jac]{Joint Astronomy Centre, 660 N.\ A`oh\=ok\=u Place, Hilo, HI 96720, USA}
\address[glasgow]{SUPA School of Physics \& Astronomy, University of Glasgow, Glasgow, G12 8QQ, UK}
\address[ohio]{Department of Astronomy, The Ohio State University, Columbus, OH 43210, USA}
\address[geminis]{Gemini Observatory, Casilla 603, La Serena, Chile}
\address[princeton]{Department of Astrophysical Sciences, Princeton University, Princeton, NJ 08544, USA}
\address[iaa]{Instituto de Astrof\'isica de Andaluc\'ia, Glorieta de la Astronom\'ia s/n, E-18008, Granada, Spain}
\address[ska]{Square Kilometre Array Organisation, Jodrell Bank Observatory, Lower Withington, Macclesfield SK11~9DL, UK}
\address[ipac]{Infrared Processing and Analysis Center, Caltech, Pasadena, CA 91125, USA}
\address[icrar]{International Centre for Radio Astronomy Research, M468, 35 Stirling Hwy, Crawley, Perth WA 6009, Australia}
\address[microsoft]{Microsoft Research, 14820 NE 36th Street, Redmond, WA 98052, USA}
\address[sao]{Smithsonian Astrophysical Observatory, 60 Garden Street, Cambridge,
MA 02138, USA}
\address[nrao]{National Radio Astronomy Observatory, 520 Edgemont Road,
Charlottesville, VA 22903, USA}
\address[aifa]{Argelander-Institut f\"{u}r Astronomie, Universit\"{a}t Bonn, Auf dem H\"{u}gel 71, 53121 Bonn, Germany}
\address[uwaterloo]{Department of Physics, University of Waterloo, Waterloo, ON N2L~3G1, Canada}
\address[warwick]{Department of Physics, University of Warwick, Coventry CV4 7AL, UK}
\address[aip]{Leibniz-Institut f\"{u}r Astrophysik Potsdam (AIP), An der Sternwarte 16, 14482 Potsdam, Germany}
\address[ucm]{Departamento de Astrof\'{i}sica, Universidad Complutense de Madrid, 28040, Madrid, Spain}
\address[unsw]{Department of Astrophysics, School of Physics,
  University of New South Wales, Sydney, NSW 2052, Australia}
\address[mpia]{Max-Planck-Institut f\"{u}r Astronomie, K\"{o}nigstuhl 17, 69117 Heidelberg, Germany}
\address[iasf]{Institute for Space Astrophysics and Cosmic Physics, Via Piero Gobetti 101, Bologna, I-40129, Italy}
\address[psu]{Dept.\ of Astronomy and Astrophysics, The Pennsylvania
  State University, 525 Davey Lab, University Park, PA 16802, USA}
\address[mps]{Max-Planck-Institut f\"{u}r Sonnensystemforschung,
  Justus-von-Liebig-Weg 3, 37077 G\"{o}ttingen, Germany}
\address[msum]{Department of Physics and Astronomy, Minnesota State University Moorhead, 1104 7th Ave. S., Moorhead, MN 56563, USA}
\address[cral]{Centre de Recherche Astrophysique de Lyon, UMR 5574, CNRS,
  Universit\'{e} de Lyon, ENS Lyon, 
  46 All\'{e}e d'Italie, 69364 Lyon Cedex 07, France}

\begin{abstract}

The Flexible Image Transport System (FITS) standard has been a great
boon to astronomy, allowing observatories, scientists and the public
to exchange astronomical information easily. The FITS standard,
however, is showing its age. Developed in the late 1970s, the FITS
authors made a number of implementation choices that, while common at
the time, are now seen to limit its utility with modern data. The
authors of the FITS standard could not anticipate the challenges which
we are facing today in astronomical computing. Difficulties we now
face include, but are not limited to, addressing the need to
handle an expanded range of specialized data product types (data
models), being more conducive to the networked exchange and storage of
data, handling very large datasets, and capturing
significantly more complex metadata and data relationships.

There are members of the community today who find some or all of
these limitations unworkable, and have decided to move ahead with
storing data in other formats. If this fragmentation continues, we
risk abandoning the advantages of broad interoperability, and ready
archivability, that the FITS format provides for astronomy.
In this paper we detail some
selected important problems which exist within the FITS standard
today. These problems may provide insight into deeper underlying
issues which reside in the format and we provide a discussion of
some lessons learned. It is not our intention here to prescribe specific remedies to
these issues; rather, it is to call attention of the FITS and
greater astronomical computing communities to these problems in the
hope that it will spur action to address them.

\end{abstract}


\end{frontmatter}


\newcommand{\ascl}[1]{\href{http://www.ascl.net/#1}{ascl:#1}}
\newcommand{\aspconf}{ASP Conf.\ Ser}
\newcommand{\aap}{A\&A}
\newcommand{\aaps}{A\&AS}
\newcommand{\jrasc}{JRASC}
\newcommand{\qjras}{QJRAS}
\newcommand{\mnras}{MNRAS}
\newcommand{\pasp}{PASP}
\newcommand{\pasa}{PASA}
\newcommand{\apjs}{ApJS}

\section{Introduction}

The Flexible Image Transport System standard (FITS;
\citealt{1979ipia.coll..445W,1980SPIE..264..298G,1981A&AS...44..363W,1981A&AS...44..371G} and
\citealt{2001A&A...376..359H}; and more recently, the definition of the
version 3.0 FITS standard by \citealt{2010A&A...524A..42P}) has been a
fundamental part of astronomical computing for a significant part of the
past four decades. The FITS format became the central means to store and
exchange astronomical data, and because of hard work by the FITS
community it has become a relatively easy exercise for application
writers, archivists, and end user scientists to interchange data and
work productively on many computational astronomy problems. The success
of FITS is such that it has even spread to other domains such as medical
imaging and digitizing manuscripts in the Vatican Library
\citep{2006JRASC.100..242W,2012EWASSAlle}.

Although there have been some significant changes, the FITS standard
has evolved very slowly since its genesis in the late 1970s. New types
of metadata conventions such as World Coordinate System
\citep[WCS;][]{2002A&A...395.1061G,2002A&A...395.1077C,2006A&A...446..747G}
representation and data serializations such as variable length binary
tables \citep{1995A&AS..113..159C} have been added. Nevertheless,
these changes have not been sufficient to match the greater evolution
in astronomical research over the same period of time.

Astronomical research now goes beyond the paradigm of a set of
observational data being analyzed only by the scientific team who
proposed or collected it. The community routinely combines original
observations, theoretical calculations, observations from others, and
data from archives on the internet in order to obtain better and wider
ranging scientific results. A wide variety of research projects now involve many
diverse datasets from a range of sources. Instruments in astronomy
now produce several orders of magnitude larger datasets than were common
at the time FITS was born, in some cases requiring parallelized,
distributed storage systems to provide adequate data rates
\citep{2012ASPC..461..283A}.

Astronomers have increasingly come to rely on others to write software
programs to help process and analyze their data. Common libraries, analysis
environments, pipeline processed data, applications and services
provided by third parties form a crucial foundation for many
astronomers' toolboxes. All of this requires that the interchange of
data between different tools needs to be as automated as possible, and
that complex data models and metadata used in processing are
maintained and understood through the interchange.

These changes in research practices pose new challenges for the
21\textsuperscript{st} century. We must address the need to handle an
expanded range of specialized data product types and models, be more
conducive to the distributed exchange and storage of data, handle very
large datasets and provide a means to capture significantly more complex
metadata and data relationships.

A summary of these significant problems within the FITS standard was
presented in \citet{P90_adassxxiii}.  Already some of these limitations
have caused members of the community to seek more capable storage
formats, both in the past, such as the Starlink Hierarchical Data System
\citep[HDS;][]{1982QJRAS..23..485D,2015HDS}, the eXtensible Data Format
\citep[XDF;][]{2001ASPC..238..217S}, FITSML \citep{2001ASPC..238..487T}
and HDX \citep{2003ASPC..295..221G}; and in the present and future
(e.g., HDF5 \citep{2011ASPC..442...53A} and NDF \citep[][\ascl{1411.023}]{2015Jenness}).
There are other popular file formats among the
radio and (sub-)millimeter astronomy community such as the Continuum and
Line Analysis Single-dish Software (CLASS) data format associated with
the Grenoble Image and Line Data Analysis Software (GILDAS) tools
(\ascl{1305.010}). Although this file
format does not have a public specification,  there are open-source
spectroscopic software packages like \texttt{PySpecKit}
(\ascl{1109.001}) that support certain
versions of the data format.  Given the large amount of available
storage formats, there is certainly a possibility that the use of FITS
will fall in favor of other scientific data formats should it not adapt
to these new challenges.

The strengths of FITS are well known and include an easily understood
serialization, a plethora of stable supporting software, good documentation of
the format and the simple fact that it remains to this day the \emph{lingua
franca} of astronomical data format exchange. What we feel has been missing
is an attempt within the community to
dispassionately discuss and understand FITS in terms of problems in
its application to modern astronomical research. In this paper we hope to show
that technologies and research techniques in astronomy have evolved but FITS has
not kept pace. As a result, gaps between FITS utility and the needs of the
research community have opened up and widened over time.
It is our intended goal in this paper to highlight some selected, important,
problems which exist in the FITS core standard today. We have deliberately
avoided proposing solutions to the problems we discuss, and we remain agnostic
(because the authors are divided) on whether replacing FITS is an obviously good
or an obviously bad idea.

We present our argument in the following manner. The various issues have been grouped
under the general topics ``information interchange'' (section~\ref{section_poor_exchange}),
``data models'' (section~\ref{section_crit_data_models}),
``metadata and data representation'' (section~\ref{section_inflex_represent}) and
``large and/or distributed datasets'' (section~\ref{section_poor_large_data_support}).
We address each of these topics in turn below then try to provide an
analysis of any deeper causes or ``lessons learned'' (section~\ref{sec:discussion}).
A summary section ends the paper and provides an overview of our work and future
direction.

\section{Information Interchange}
\label{section_poor_exchange}

FITS originated as a delivery format for observatory data. It was the format
of choice when transporting data between different data reduction
environments such as IRAF (\ascl{9911.002}),
Starlink (\ascl{1110.012}), AIPS
(\ascl{9911.003}) and MIDAS
(\ascl{1302.017}).

In principle, FITS promotes interchange through its simple and easily
understood format which holds its information in various levels of groupings
of metadata and data blocks. Metadata are captured via key-value pairs which
are in turn grouped into FITS headers. The first header is denoted as the
`primary' header and subsequent headers known as `extensions'. Headers may or
may not be then grouped with data blocks.  An example primary FITS header
appears in Fig.~\ref{fig:fitshead}.

\begin{figure*}
\begin{minipage}{\textwidth}
\begin{lstlisting}
SIMPLE  =                    T / Standard FITS format
BITPIX  =                  -32 / 32 bit IEEE floating point numbers
NAXIS   =                    3 / Number of axes
NAXIS1  =                  800 /
NAXIS2  =                  800 /
NAXIS3  =                    4 /
EXTEND  =                    T / There may be standard extensions
ATODGAIN=             7.000000 / Analog to Digital Gain (Electrons/DN)
RNOISE  =             1.010153 / Readout Noise (DN)
EPOCH   =       49740.82869315 / exposure average time (Modified Julian Date)
EXPTIME =          2500.000000 / exposure duration (seconds)--calculated
EXP0    =          1300.000000 / weighted average initial exposure time
RSDPFILL=                 -250 / bad data fill value for calibrated images
SATURATE=                10237 / Data value at which saturation occurs
TEMP    =                    0 / Temperature (0=cold, 1=warm)
FILTNAM1= 'F555W             ' / first filter name
HSTPHOT =                    T / Preprocessed by HSTphot/mask
END
\end{lstlisting}
\caption{Representative simple primary header of a FITS file showing
  an assortment of FITS keywords and their associated values. This
  header from 1995 uses a definition of the, now deprecated,
  \texttt{EPOCH} keyword that is at odds with the standard usage of
  the period but the lack of parsable units for the field make it
  hard for a computer parser to understand this.
  Bytes which contain data may or may not follow the \texttt{END} keyword of
  the header.}
\label{fig:fitshead}
\end{minipage}
\end{figure*}

This simple arrangement of information can satisfy many use cases for
transport, however, requirements for interchange have evolved. Effective
interchange, as we shall illustrate, now includes things like the ability to
declare models for use in higher level processing, validation of models within
the file and, at the most basic level, the ability to declare which version of
the serialization is being used.

These capabilities have been explored and implemented in several other data formats in astronomy.
The Astronomical Data Center (ADC) XDF
format, the Low-Frequency Array for Radio Astronomy (LOFAR) HDF5 data
model \citep{2012ASPC..461..283A}, CASA measurement sets
\citep{2012ASPC..461..849P},
RPFITS\footnote{\url{http://www.atnf.csiro.au/computing/software/rpfits.html}
-- RPFITS is an incompatible fork of FITS \citep[see e.g.,][]{1998ASPC..145...32B}.} from the Australian Telescope
National Facility
and Starlink's NDF
\citep{1988STARB...2...11C,1993ASPC...52..229W,P91_adassxxiii} all
serve as examples in this regard.

XDF was created primarily to support archiving, web-based use of
published astronomical data and the development of FITSML -- an XML version
of the FITS data model which could use an XML schema for validation.
NDF was developed in the late
1980s as a means of organizing the hierarchical structures that were
available via the Starlink HDS format when it became apparent that
arbitrary hierarchies could lead to chaos and lack of ability for
applications to interoperate \citep{2015Jenness}.
HDX \citep{2003ASPC..295..221G} was developed around 2002 as a flexible
way of layering high-level data structures, presented as a virtual XML
Document Object Model (DOM), atop otherwise unstructured external data stores; this was in
turn used to develop Starlink's NDX framework, which (among other
things) allowed FITS files to be viewed and manipulated using the
concepts of the NDF format.
HDF5 \citep{2012ASPC..461..283A} was chosen to accommodate LOFAR's
exceptional high data rates, 6-dimensional data complexity, distributed
data processing and I/O parallelization needs.

\subsection{Format versioning}
\label{subsection_format_versioning}

There is no standard means for a FITS file to communicate
the formatting version it conforms to.  Consider the example primary
header in Fig.~\ref{fig:fitshead}: the only keyword which implies any
type of format is SIMPLE which is set to `T', or true. The comment
indicates that the file conforms to ``Standard FITS format'', but what
indeed is that `Standard'?

The designers and maintainers of FITS have espoused the principle
``once FITS, forever FITS''
\citep[see e.g.,][]{1988A&AS...73..359G,1993FITS1}.  Certainly some in the
community see this as a strength for the format as it appears to
promote long term stability and ``archivability'' of FITS data
\citep{2012EWASSAlle,2012LOC}.  This is not, however, quite the same
thing as saying that FITS is unversioned.  There have been at least
three named descriptions of FITS.  These include the first, or `basic
FITS' document \citep{1979ipia.coll..445W,1981A&AS...44..363W}, the
NOST version of FITS \citep{2001A&A...376..359H}, and the current
version 3.0 \citep{2010A&A...524A..42P}.  One can regard these as
successive improvements of a document describing changing best
practices for an unchanging format (compare ``the value [of the
putative FITS version keyword] is always 1.0 by default'' in
\citet{1997ASPC..125..257W}, which discusses this general point in
some depth). However the fact remains that there are features in the
most recent FITS description (such as 64-bit integers, negative
\texttt{BITPIX} values, FITS extensions and tables) which were not
present in the first FITS version and demonstrably FITS has evolved.

The ``once FITS, forever FITS'' doctrine may be taken to require
backward and forward compatibility or, if you will, compatibility
with all FITS files ever created in the case where there is only
one version ever.
Either way, backward compatibility means that it always
\emph{should} be feasible to use the most recent FITS reader. For
forward compatibility, at minimum, reasonable expectation goes
beyond requiring a FITS reader not crash when confronted with a newer
FITS file; it should do more than this. Ideally, it should parse
what parts of the file are still compliant with
its understanding of the format and report on those parts/features
of the file which it does not recognize. In either compatibility case,
without unambiguous version metadata, readers have to rely on `duck-typing'
\footnote{see \url{http://en.wikipedia.org/wiki/Duck\_typing}}
and heuristics which are ultimately error prone because it requires
the implementer of the parser to perfectly interpret the signature
of any particular set of features present in the given FITS instance
from among other possible features which are absent. Furthermore, as
the format evolves beyond the date of its creation, the software cannot
know how that signature may change and may incorrectly
identify the version, a clear difficulty for forward compatibility.
The reliance on heuristics also has impact beyond writing a FITS parser.
Future archivists will certainly want to know what version of the
format they are dealing with without having to guess from ancillary
evidence such as the presence of certain keywords, date of the file
creation and so on.

The lack of versioning also limits the ability of our community to
move forward constructively with developing new FITS versions.
The ``once FITS, forever FITS'' doctrine requires we accrete
more and more ``design rules'' which may limit our ability to implement
new and needed features and clutter reader code. Consider that three keywords
have been deprecated
(\texttt{BLOCKED}, \texttt{CROTA2} and \texttt{EPOCH}) by the latest version
of FITS. Per the standard, these are ``obsolete structures that should not be
used in new FITS files but which shall remain valid indefinitely''.
As such, software writers must indefinitely be on guard for these metadata
and writers of new conventions must avoid utilizing these specific keywords.
As time passes and changes of this nature accumulate, it will be progressively
harder to interpret FITS data correctly and write new conventions.

Although the FITS format is apparently rather simple, on disk, the
multiple versions of the format description, and the existence of
numerous header conventions, mean that reading a FITS file in full
generality is a complicated and messy business.  As there is no
versioning mechanism to effectively declare deprecated structures
finally ``illegal'', these complications and costs will only
increase.

\subsection{Declaration and validation of content meaning}
\label{subsection_semantic_validation_and_declaration}

Related to, but separate from, the lack of versioning of the
serialization, is the lack of ability to declare the presence of data
models and their associated meaning.  By `data model' we
mean:

\begin{quote}
``a description of the objects represented by a computer system
together with their properties and relationships; these are typically
`real world' objects such as products, suppliers, customers, and
orders.\footnote{Definition adopted from Wikipedia, see
\url{http://en.wikipedia.org/wiki/Data\_model}}
''
\end{quote}

Of course, objects in astronomy are more likely to involve things like
observations, instruments, celestial coordinates and actual astronomical
objects such as stars. Likely properties one will encounter in a FITS
file include things like observational parameters (start/end times),
astronomical coordinates, name and properties of the observing
instrumentation, and so forth. In FITS-speak, we can say that any FITS
keyword outside those defined in the FITS standard is a data model
parameter, and collections of related FITS keywords form a data model.
Ideally a data model should be associated with a given, unique,
``name\-space'' so that collisions in naming of the models and requisite
parameters are avoided.

Data models can provide a standard by which information (data and metadata)
in the file may be semantically and syntactically validated in software.
Questions such as ``are all of the required metadata/data structures present
in the file?'' (e.g., all of
the needed keywords occur in the correct places in the file) and ``are
there any non-normative values in the file?'' (all metadata/data values
are within expected bounds) are both questions answered by syntactic
validation, the conformance of information in the file to one or more
declared data models.  The question of ``how do these data (inter)relate
with other data'' (e.g., can named structures in the file be associated
in some manner with others in another file/extension?) is one of semantic
validation. By confirming that the file is `valid' in both senses, we may
link the data model to the information in the file, and hence answer the
fundamental question ``what does this data you gave me represent?'' (e.g., lists
of stars, tables of galaxies, images of dust clouds, etc).
It is important to note that all of these questions are critical to
consumers of the file.

There is already evidence that the FITS community values and needs shared
data models. There are many examples. WCS and some other FITS conventions
such as OIFITS
\citep{2006SPIE.6268E.106T}, MBFITS \citep{2006A&A...454L..25M},
PSRFITS \citep{2004PASA...21..302H},
SDFITS \citep{2000ASPC..216..243G} and FITS-IDI \citep{2011AIPS114}
are data models. The declaration of keyword
dictionaries\footnote{Some collected data dictionaries with FITS
keywords may be seen at the GSFC FITS site, see
\url{http://fits.gsfc.nasa.gov/fits\_dictionary.html}} is also essentially
an act of declaring one or more data model(s).

Let us also note that it is not unreasonable to expect more than one model
to appear within a file. Consider data distributed by the Palomar Transient
Factory.  For these data to permit the widest
variety of software tools to understand the astrometric distortion in these
images, keywords from both the ``SIP'' and ``TPV'' conventions are included
\citep{2012SPIE.8451E..1MS}.
One convention expresses distortion polynomials in pixel space and the
other in intermediate longitude and latitude, yet it is not immediately
obvious which data model should be applied.

All of these data models imply an associated ``namespace'' which is
a means of declaring the origin of the data model so that we may
disambiguate and/or associate declared properties between models.
For example, separate namespaces should exist for the two aforementioned
astrometric distortion models in the example above.
There are common problems which name\-spaced models help to solve and even
the `simple' metadata in Fig.~\ref{fig:fitshead} illustrates this.

Consider the \texttt{TEMP} keyword in the example. Without reading the comment
associated with it, we cannot know if this is this a temperature or perhaps
some type of temporary file or resource or something else. If it is a
temperature then what is this the temperature of? What do the values `0'
and `1' mean? Are these the only valid values for this keyword?  \texttt{TEMP} is
a likely keyword string to appear in other files, how do we know if the
\texttt{TEMP} in the other files is the same one we see in the example?

Clearly, it is a non-trivial matter for the machine to determine whether
these are the same properties and to know other important details for using
this information. This problem is not isolated to a solitary bit
of rogue metadata. We can ask similar questions about most of the keywords
in the example header. Namespaced data models help address these issues. With
an appropriate namespace mechanism in place, it is possible to create a
machine-readable mapping between the data models so that any software program
can determine whether \texttt{model1:TEMP} is the same (or different) property as
\texttt{model2:TEMP}.
Namespacing mechanisms can both provide humans with documentation, and
provide software with the means to look up model definitions (perhaps from
remote locations), and thus apply syntactic or semantic validation rules
for the information at hand.  This will allow the program to
answer the remainder of our posed questions above.

These arguments indicate there is a pressing need for name\-spaced data models,
yet, the only way in which we can currently implement them is for a human
to inspect the file, or to write special purpose software programs targeted to
particular data models. Given the data volumes that we have in astronomy, the
latter choice is in the direction we should go, but is not practical
in the general case.

The writing of generalized software programs to detect any data models
present in a given FITS file is currently a difficult task for many reasons.
First and foremost, we must recognize that there are constantly new data models
being created and modified. Some of these are documented in a human readable
fashion but there are many more models which do not even meet this standard.
Worse, due in part to the lack of good validation tools, the community has
accepted many informal variants of existing models. These variants may both
be documented or not but are a result of either accidental or intentional
stretching of the original metadata usage. The header in
Fig.~\ref{fig:fitshead}, for example, is an informal variant because of its non-standard
use of the \texttt{EPOCH} keyword.
Finally, there is the possible complication of more than one data model being
fully, or partially, present within a file. Without explicit signposts for
the software to use, it is likely impossible to determine which data models
are present and map information to appropriate meaning.

\section{Data models}
\label{section_crit_data_models}

One of FITS strengths is that it includes some common data structures
which are important in astronomy data. The FITS standard includes such
things like ``table'' and ``n-dimensional array'' the latter which is
used to model both images and data cubes.  These items are really simple
representations of the data at a primitive level, and are certainly
needed for basic access to the information within the file.  Even so
these structures, by themselves, do not contain much in the way of necessary
detail and semantic information which tells the consumer exactly what
it is they are actually consuming. For this reason, they cannot be considered to
be data models.

The FITS standard does supply a data model, for example the aforementioned
WCS may be considered to be part of it, and these standard semantics are generally
regarded as another strong point of FITS.  The other data formats we have
previously mentioned vary in how extensive their core data model is. The range
goes from HDF5, which does not supply any data model \emph{per se}, to that of
NDF which has very rich metadata in its data model.

It is a matter of opinion as to whether more/richer detailed data models in the
format standard are better or not. The NDF core data model metadata are certainly
more detailed than the metadata in the FITS standard. On the other hand FITS is
certainly more widely adopted than NDF. Nevertheless, we believe FITS would benefit
from an expansion in its standard data model as there are certainly common semantics
which may be found in other data formats (e.g. NDF, XDF, etc) and FITS-based model
extensions (e.g. such as MBFITS or local data dictionaries) which the community can
benefit from.

In this section we detail some important missing (component) data models.

\subsection{Scientific Errors}

The measurement of physical properties with their associated uncertainties
is fundamental to astronomical
research. It is thus ironic that FITS, which is purposely designed for
supporting astronomical research, has no standard data model for
capturing information about scientific errors.

We could easily list a great number of possible error types which
might be useful but trying to encompass all of the needs of the
community at once is likely to create an unwieldy data model. We
suggest that the community needs to provide for the most common needs,
and target that subset as a first, shared model. Earlier efforts which
might inform and help this work include local data models at sites
such as CADC \citep{2012ASPC..461..339D} and the error
models implemented in other data formats like NDF
\citep[although see for example][]{1991STARB...8...19M}, and software
efforts underway in scientific programming communities such as Astropy
\citep{2013A&A...558A..33A}. Each of these has valuable insight into
the requirements.

Nevertheless, we can anticipate that the following general
characteristics might be part of the model:

\begin{itemize}
\item Allow for both metadata and data to have errors.

\item Allow for extensible classification of the error type. For example,
``Gaussian'' errors are also a subclass of ``statistical'' errors.

\item Allow association of more than one error class/type per
measurement. For example, allow for both systematic and statistical
errors to be associated with each measurement.

\item Allow for additional properties to be associated with each error
class. For example, ``statistical'' errors may have an assigned ``sigma''
value.
\end{itemize}

\subsection{Extended Coordinate Support}
\label{sec:wcs}

The existing FITS WCS data models illustrate some of the limitations
associated with FITS. The ``once FITS, always FITS'' idea required that
the current WCS standards were developed as an extension of the older
AIPS standard, and so inherited many of the inherent limitations of
that system. Even so they took a long time to be agreed. They are
complex yet incomplete and inflexible. They are inadequate for many
modern telescopes, and restrict creative use of novel coordinate
transformations in subsequent data analysis. For instance, raw data
must handle more distortion issues than the FITS WCS standard
projections can handle. There are some provisions for handling more
arbitrary distortions, but they are either cumbersome or too
simple. Perhaps the biggest limitation is that different
transformations of coordinates cannot be combined in flexible
ways. The user is effectively limited to choosing only one of the solutions
available.

This is unfortunate. Not only does it reduce the range of
transformations that can be described, but it also makes it harder to
decompose the total transformation into its component parts thus making
understanding and manipulation of the total transformation harder. The
alternative approach -- a ``toolkit''-style system that creates complex
transformations by stacking simpler atomic mappings -- is usually the
most efficient representation as far as data storage is concerned
(for example AST, see below).

To illustrate the problem consider the imaging data taken by the
Hubble Space Telescope which require multiple distortion components \citep[see e.g.,][]{2013ASPC..475...49H}.
Some are small but discontinuous. Others are linear but time varying.
There is no FITS WCS compatible solution that handles these needs well.
As another example, SCUBA-2 raw data \citep[see
e.g.,][]{2013MNRAS.430.2513H} include focal plane distortions which are
combined with other transformations but must also support the dynamic
insertion of other distortion models when a Fourier transform
spectrometer \citep{2010SPIE.7741E..67G} is placed in the beam.

Another case with poor support is Integral Field Unit (IFU) data.
Many of these datasets
have discontinuous WCS models. The only way to support these in FITS
now is to explicitly map each pixel to the world coordinates. Besides
being space inefficient, it is difficult to manipulate in any simple
way.

In addition to limiting the description of raw telescope data, FITS
WCS also restricts what can be done with such data during subsequent
analysis. There are many potentially interesting transformations that
would result in the final WCS being inexpressible using the
restrictive FITS model. For instance, transforming an image of an
elliptical galaxy into polar or elliptical coordinates is currently not possible. Another
case which is unworkable is an alternate coordinate system to an image to
represent the pixel coordinates of a second image covering the same
part of the sky.  These may not be common requirements, but they
illustrate the wide range of transformation that should be possible
with a flexible WCS system.

The inflexibility in the FITS solution arises from multiple issues,
but lack of namespaces is a serious barrier to providing a more
flexible solution. If one has multiple model components each with
similar parameters, how does one distinguish between them? One may use
the letter suffix, but that is also used to distinguish between
alternate WCS models. The limitation on keyword sizes presents
limitations on how many coefficients can be supported. The lack of any
explicit grouping mechanism requires complex conventions on how to
relate whole sets of keywords. With more modern structures, such
contortions and limitations are not necessary.

The reality is that to solve these problems, many software systems
have chosen alternate solutions and save their WCS information in FITS
files in other ways (or in separate files). For example, the AST
library \citep{1998ASPC..145...41W,2012ASPC..461..825B} is not subject
to these limitations, but
is forced to use non-standard FITS keywords when serializing mappings
to FITS files (see Fig.~\ref{fig:asthead}).

\begin{figure*}
\begin{minipage}{\textwidth}
\begin{lstlisting}
PLRLG_A =     5.50788096462284 / Polar longitude (rad.s)
ENDAST_K= 'SphMap  '           / End of object definition
MAPB_A  = '        '           / Second component Mapping
BEGAST_O= 'CmpMap  '           / Compound Mapping
NIN_D   =                    3 / Number of input coordinates
NOUT_B  =                    2 / Number of output coordinates
INVERT_C=                    0 / Mapping not inverted
ISA_I   = 'Mapping '           / Mapping between coordinate systems
INVA_B  =                    1 / First Mapping used in inverse direction
MAPA_C  = '        '           / First component Mapping
BEGAST_P= 'MatrixMap'          / Matrix transformation
NIN_E   =                    3 / Number of input coordinates
INVERT_D=                    1 / Mapping inverted
ISA_J   = 'Mapping '           / Mapping between coordinate systems
M0_A    =    0.426766777415161 / Forward matrix value
M1_A    =    0.699933471661958 / Forward matrix value
M2_A    =    0.572680760059142 / Forward matrix value
M3_A    =   -0.418237169285184 / Forward matrix value
\end{lstlisting}
\caption{Example header of a representation of an AST WCS object in a
  FITS header  when the mapping is too complex to be represented using the
  FITS-WCS standard.}
\label{fig:asthead}
\end{minipage}
\end{figure*}

\subsection{History and Provenance}
\label{sec:history}

The FITS standard encourages people to store processing history
information in the header using a pseudo-comment field named
\texttt{HISTORY}. This works from the perspective of making the information
available to a sufficiently interested human (assuming that each step in the
data processing adds information to the end of the history section of the
header) but the free-form nature of the entries makes it essentially
impossible for a software system to understand what was done to the data.
This may be possible within the constraints of a single data reduction
environment but it is highly unlikely that the content of the
\texttt{HISTORY} block can be understood by any other software packages. History
needs to be treated as a first-class citizen with a standardized way of
registering important information such as the date, the software tool and
any relevant arguments or switches.

A related issue is data provenance; that is, sufficient records of how files
were created to permit their reproduction. For a given processed data product
it is, for example, impossible to determine which data files
contributed to the creation of that product. While there is no metadata standard
for specifying this
information in output files, experimental systems have been developed which, when fully developed,
aim to offer programmatic interfaces that will
simplify recording provenance information. One such example is
Provenance Aware Service Oriented Architecture
\citep[PASOA;][]{2008IPAWMoreau,2011743Moreau}, an open source architecture
already used in fields such as aerospace engineering. In brief, when
applications are executed they produce documentation of the process recorded
in a repository of provenance records that are persisted in a database. In
astronomy, PASOA was successfully demonstrated by integrating it into the
Pegasus workflow management system for running the Montage mosaic engine
\citep{2009SCGroth}.

At the JCMT Science Archive \citep[JSA;][]{2008ASPC..394..135G,2015Economou} data are
created with full provenance information using the native provenance
tracking that is part of NDF \citep{2009ASPC..411..418J}. This
provenance includes every ancestor along with history information that
contributed to each ancestor. When these files are converted to FITS
for ingestion into the JSA using the CAOM-2 data model
\citep{2013ASPC..475..159R} the provenance is trimmed to include
just the immediate parent files (using \texttt{PRVnnnnn} headers)
and the observation identifiers of the root ancestor observations
(using \texttt{OBSnnnnn} headers). The full richness of the provenance
information is available in FITS binary tables but the lack of a standard
leaves this information hidden from applications other than the ones
that created it originally.

Finally, astronomy may benefit from methodologies used to develop provenance
systems custom to Earth Science and remote sensing
\citep{2008IPAWTilmes,2008IPAWMcCann}.

\subsection{Data Quality}

One of the more pressing needs in our era of shared and distributed
data is the need to know which data are ``good'' or, to put it another
way, of sufficient quality. We are long past the era when the data
volume was so small that it is practical to download all of the possible
data of interest and examine it locally.

Some might insist that this is an easily solved problem. Simply
declare a keyword, like \texttt{DQUALITY}, and allow it to take a boolean
value. To be sure, that example is an exaggeration, but it helps to
illustrate that there is no single optimum between the virtue of
simplicity and the vice of being simplistic.
Data quality cannot be judged on a single, or even a
small set, of parameters. The
data which are adequate for one type of use, may be wholly inadequate
in another usage context. Consider that engineering data generally are
unsuitable for science and vice versa. Science data may be
unsuitable for other types of science (for example, studies of sky
background vs.\ pointed source science).

A data quality model then, should be an ensemble of common statistical
measures of the type of dataset which may be used to derive
higher-level judgments of the quality/suitability of the data for
some other declared purpose. There are many higher types of data
quality models which will need be created from the lower-level
measures (image data quality, pointed catalog data quality, etc) and
from these particular, targeted, statistical measures data quality may
be judged by the dataset consumer without directly examining the data
themselves.

\subsection{Units}
\label{sec_units}

A strength of FITS is that it includes support for units within its core
standard. There are, however, limitations in the utility of the provided
specification.

First, while it syntactically flexible, there are a few
specification ambiguities which could be resolved by an explicit grammar.
This limitation has perhaps been one reason that others
have felt the need to publish more explicit prescriptions for units
\citep{1995OGIPUnits}.  Another limitation is that the model does
not accommodate the full range of contemporary astronomical data. This
is evident from the adoption of other units systems by some major
archives such as the CDS (Centre de Donn\'ees astronomiques de Strasbourg\footnote{\url{http://cds.u-strasbg.fr/}}) which contains a large number of published
astronomical tables (see the units section within \citealt{2000CDSUnits}).
Finally, there is also a provenance issue to defining new units, since
-- to pick the example of
the unit of `Jupiter radius' -- two different groups may prefer the
mean or equatorial values for the radius, or in contrast may regard it,
not as simply an abbreviation for a certain number of
kilometers, but instead as a distance whose value is determined at a
certain atmospheric pressure level.

The solution to these limitations is not simply to expand the list of
recommended units since, as well as being slow, this fails to distinguish
between, for example, different definitions of the second, or to communicate
places where the distinction does or does not matter.

The purely syntactical issues surrounding unit strings are
being addressed by the IVOA's `VOUnits' work \citep{VOUnits}, but the higher
level questions -- of communicating and defining new units, of
indicating documentation, and of converting between them in a
scientifically meaningful manner -- are out of scope for that work by
design, since experience has shown them to be more contentious than
one might expect. These questions should be taken up by the FITS community.

Solutions then may involve cherry-picking VOUnits syntactical fixes and
alteration of the units model to use some namespacing mechanism which
would help to disambiguate sources of extended units models found within a file.
Finally, we believe that an analysis of other more recent work, such as that
found in \citeauthor{VOUnits} and \citeauthor{2000CDSUnits} can help to quickly
round out the roster of standard units.

\section{Metadata and data representation}
\label{section_inflex_represent}

What may be construed to be ``FITS data'' has changed significantly
since the founding of FITS. The original FITS specification mandated
only the capture of astronomical images. Almost a decade later, FITS
extensions \citep{1988A&AS...73..359G}, allowing for gathering
multiple related data structures in one FITS file, and ASCII tables
\citep{1988A&AS...73..365H} were introduced. Binary tables followed in
the next decade \citep{1995A&AS..113..159C}. Changes have also
occurred in metadata capture. Over the intervening years the FITS
community has added new metadata conventions such as \texttt{HIERARCH}
\citep{2009Wic} and \texttt{GROUPING} \citep{2007Jen,1995ASPC...77..229J}, which have
allowed for greater flexibility in capturing metadata.
This expansion in capability to serialize information is to be
expected and is all to the good. Nevertheless, as we shall illustrate
below, the expansion is still insufficient relative to actual need.

\subsection{Rich metadata representation}
\label{subsection_information_representation}

The basic element of metadata capture in a FITS file is the FITS
``card'' which comprises a keyword name, a keyword value and a
comment. All FITS cards must contain the keyword name and keyword
value pair while comments in cards are optional. Comments may
sometimes contain information about units of the metadata value.

Metadata may comprise a rich assortment of data structures and
single-valued metadata are only the beginning. There is a need to
capture sets, lists, vectors and objects within metadata (to name only
the most basic of structures). Yet FITS cards, without additional
conventions, are only capable of capturing single, scalar keyword-value
pairings.

The expression of both objects and non-scalar, multi-valued keywords
is difficult in FITS and data model designers have to resort to
conventions to achieve this. Object storage is enabled in part by
utilizing a hierarchical convention such as \texttt{HIERARCH} or the
\emph{record-valued} system proposed in the FITS distortion paper
\citep{FITSDistort}.  In order to hold a keyword
with either a `set' or `list' value, a common local convention adopted is
to create a set of keywords sharing the same base name followed by a
integer value which may (or may not) indicate order of the values
(such as \texttt{ICMB001}, \texttt{ICMB002}, and so on). Another example
is the IRAF \textit{multispec} format
\citep[see][and references therein]{1993ASPC...52..467V} which uses
this scheme to specify related world coordinate information (see
Fig.~\ref{fig:multispec} for an example).  The AST library
\citep[][and see also \S\ref{sec:wcs}]{1998ASPC..145...41W} takes a
similar approach in converting the WCS objects into FITS headers when
the transformations are too complex to be represented by standard WCS
headers (see Fig.~\ref{fig:asthead}).

\begin{figure*}
\begin{minipage}{\textwidth}
\begin{lstlisting}
WAT0_001= 'system=multispec'
WAT1_001= 'wtype=multispec label=Wavelength units=Angstroms'
WAT2_001= 'wtype=multispec spec1 = "1 113 2 4955.4428886353510.055675655'
WAT2_002= '83 256 0. 23.22 31.27 1. 0. 2 4 1. 256. 4963.0163112090 5.676'
WAT2_003= '976664 -0.3191636898579552 -0.8169352858733255" spec2 ="2 112'
WAT2_004= '9.09" spec 3 = "3 111 2 5043.5"                              '
\end{lstlisting}
\caption{Example header from an IRAF \textit{multispec} dataset
  indicating the use of multi-line headers that differs from the
  \texttt{CONTINUE} convention.}
\label{fig:multispec}
\end{minipage}
\end{figure*}

There are also restrictions on the expression
of scalar values in headers. Consider that FITS cards are limited to 80
characters and FITS keyword names may be no longer than 8 characters. The
result of these constraints is that keyword values may be no longer than
68 characters. Of course, if you use all of the space for keyword values,
then the comment, or keyword values longer than 68 characters will need
another convention in order to capture it (such as creating a continuation
line in the header using the \texttt{CONTINUE} convention \citep{2007Continue}).

Let us now consider the impact of keyword name constraints. Not only
are keyword names limited to a small set of characters but keyword
names are restricted to no more than 8 characters. Often these
restrictions prevent clear labelling of the metadata element because
authors are forced to map longer, more descriptive, names into the
truncated size. Non-English speaking authors are additionally forced to map
into the limited character set. If you doubt this leads to problems,
try the following experiment: open any non-trivial FITS file and scan
the header. Unless you are an expert in the data models present in the
file (and sometimes even if you are) it is easy to find that the
cramped names of the keywords often leads to arcane and confusing
metadata.


These restrictions on the FITS card have impact on conventions with
resulting limits on the utility of any implementation. Due to the limited
namespace and size of the keywords, different conventions often reuse
the same keywords for different purposes.  For example, compare the use
of \texttt{PV} keywords in the products of the SCAMP tools
\citep{2006ASPC..351..112B}, used for polynomial distortion coefficients,
to the more common \texttt{PV} keywords used in the WCS convention for generic
parameter values.
These two conventions, when used in the same file, cause ambiguity and
incorrect representation of the data. It is true that FITS libraries
will often provide a means to choose between one duplicate keyword or another
so that in the strictest terms the issue may be resolved.
Nevertheless, there is no guarantee that the author of
the files intended resolution per the manner of library in use, nor is there any
guarantee that different libraries will resolve the matter in the same way
so the same file may wind up with different meaning if read by different parsers.
Finally, even if libraries were consistent in behavior, the original intention
of the author is still ambiguous (for example, they may not want to resolve
in favor of one model or the other and both models are important to keep).

\subsection{Expanded missing values support}

Missing values are a common feature of most datasets, and are distinct from
invalid values (such as \texttt{NaN} or \textit{Not a Number}) that may occur
for example in floating point calculations. For images with integer data
types, one can make use of the \texttt{BLANK} keyword to represent missing
values, and for tables with integer and string columns, one can make use of
the \texttt{TNULL} header keyword. However, for floating point images or
table columns, there is no mechanism for specifying missing values. This has
led to the common use of \texttt{NaN} to represent missing floating point
values. However, one should carefully distinguish between true missing values
(which in an image could indicate for example an area of sky that was not
observed), versus an invalid value (represented by \texttt{NaN}) which may
represent for example a saturated pixel; such a distinction is not currently possible
in FITS.

\subsection{Data associations}
\label{section_data_associations}

As data acquisition and data reduction systems have become more
complex there has been a move to storing multiple image data
components in extensions within a single FITS file. The FITS extension
mechanism provides a scheme for having multiple images but, as noted
in \citet{2003ASSL..285...71G}, in essentially a flat structure
without hierarchy or inheritance. If you have nine images in the file
there is no way of indicating that three of them are data, three are an error
and three are a quality mask. Indeed, there is no way of specifying which
triplets are related. You can use the \texttt{EXTNAME} header to
indicate relationships but this relies on convention and string
parsing rather than being a standard part of the format.

As a real world example of this problem, consider the data processing
system for the Herschel Space Observatory which includes
context products that serve as containers for groups of data products
with each product capable of being mapped to a
FITS file stored on disk
\citep[see the Herschel architecture and design document;][]{2008HerschelDesign}.
In particular, Herschel's observational data
hierarchy allows all products associated with an observation (telemetry,
calibration, raw and processed data) to be linked with the capability of
lazy loading of products from the archive ``cloud''.
Satisfying the requirement that all products are storable as FITS files
has forced the links in these hierarchies to be specified in a very
convoluted form, understandable only within the
Herschel interactive processing environment
\citep[HIPE;][]{2010ASPC..434..139O} and not by other FITS readers.

Another approach to this situation is the conversion of NDF
format files to FITS and back to NDF \citep{SUN55,1997STARB..19...14C}.
They demonstrated that you can represent a hierarchical data grouping
in the FITS multi-extension format, but this is done using \texttt{EXTNAME}
conventions combined with headers representing the extension level in
the hierarchy and the type of component and so is not understood by
other FITS tools.

In both cases above, a standardized way of specifying relationships
between extensions would be extremely valuable to data and application interoperability.

\subsection{Declaring byte order}
\label{section_byte_order}

The original FITS standard specification \citep{1981A&AS...44..363W}
requires that a series of consecutive bytes in multi-byte data items is
stored in order of decreasing significance (known as big endian format).
Sometimes the byte order needs to be checked and swapped to the opposite
byte ordering (little endian format) in systems that do not support
non-native data formats.  This is the case in some implementations of FITS
readers that do not use the \textsc{cfitsio} library
(\ascl{1010.001}) and which use C routines
to implement other scientific capabilities.  Programmers on little endian
platforms who work with large data volumes may find that this limitation
results in a performance penalty as marshaling data to and from
the FITS big endian ordering will be required.  This is a frequent problem
for astronomical programs. Little endianness is found on x86 and x86-64
processors that are commonly used in universities and research
laboratories.

The inability to specify the byte order will obviously result in a
need to byte swap data. In most cases, this is not a significant
problem or impact on performance for modern software systems and
can be discounted. There is however, another, more significant issue
tied to this limitation. The ability to wrap/translate existing data products
into FITS files, without reprocessing them to the specified
byte-order in the FITS standard, is important. From
the perspective of an archivist with the responsibility of preserving the
records of astronomical observations, the less the data are altered, the
more efficient and reliable the archival data management will be.

\subsection{Alternative encodings}
\label{section_char_encoding}

The allowed character set for metadata and data in FITS is overly
restrictive and is limiting its application. The restrictions between
metadata and data do not differ significantly. For metadata,
FITS only supports the 7-bit
ASCII encoding for keyword values and comments. For data encoding,
authors may again only use 7-bit ASCII for text (string) capture in
either ASCII or binary FITS tables although the NULL character is
allowed in certain cases in binary tables.

The world of astronomy has evolved beyond capturing of scientific
information in 7-bit ASCII encoding. The FITS community has grown.
Many data are captured by instruments designed, built and run by
investigators based in non-English speaking countries and astronomical
research has grown significantly elsewhere in the world. Whether as
original observational data products, reduced data, information from new
services or in capturing theoretical data, FITS is now required to hold
data which are not exclusively originating in English-speaking countries.

The current restrictive character set is an
anachronism, particularly considering that the most common language on the planet,
Mandarin, cannot be used easily in a FITS file. Forcing information
into English can easily result in loss of valuable meaning, unnecessarily
limit the audience who may use the file, or force the author to use some other
format to store their data.

Support for alternative encodings is needed.
Simple issues which revolve around the value of keywords like the
expression of a person's name (with accents, for example), or the
ability to use special scientific and mathematical symbols (like the
\r{a}ngstr\"{o}m symbol \r{A} or the degree symbol \degree) should be
handled. Tabular text values should similarly be allowed alternative
encodings for the same reasons.  Furthermore, while not as critical, the
format should also allow keywords themselves to be expressed in a broader
range of characters.

\section{Large or distributed datasets}
\label{section_poor_large_data_support}

At the time when FITS was developed, the primary media used for
archiving and transporting the data were tapes. A magnetic tape is
unlike a hard-drive in that it is a serial access device.  The concept
of sequentially accessing the data was naturally adopted for the FITS data
model.  Although tapes are still widely used for archiving data,
such an access mode is no longer commonly available as the files are usually
transferred to a hard-drive before being accessed. What is more, the
serial nature of FITS has become a significant bottleneck when it comes
to working with large datasets.

Consider that many new instruments, especially in radio astronomy
(ASKAP;~\citealp{2009IEEEP..97.1507D},
MWA;~\citealp{2013PASA...30....7T}, LOFAR;~\citealp{2013A&A...556A...2V}, and
SKA;~\citealp{ska-exascale})
have been producing, or are planning to produce in the near future,
spectral-imaging data-cubes of unprecedented volumes in the order of
tens and hundreds of petabytes per year. Due to the increased spatial and frequency
resolutions there are individual datasets which can now be expected to be as
large as tens of terabytes.

For many reasons which we will detail below, FITS does not provide sufficient
support for these types of large data.

\subsection{Parallel write/read operations}
\label{subsection_parallel_io}

Large datasets require parallel read/write operations to be processed on
parallel computers.  FITS, however, cannot support optimization for parallel
read/write operations.

This has been the driving factor for LOFAR to
invest a significant effort into development of a new format using
HDF5  \citep{2012ASPC..461..283A}. Most of LOFAR's standard data products
are now stored using the HDF5 format, as well as HDF5 analogs for traditional radio data
structures such as visibility data and spectral image cubes.  The HDF5 libraries allow
for the construction of distributed files and impose no limits on their sizes.
The nature of the HDF5
format further provides the ability to custom design a data encapsulation format,
specifying hierarchies, content and attributes.

\subsection{Streaming imaging data}
\label{subsection_stream_image}

While FITS supports cutout capabilities, serving large datasets to an
end user requires support for multiple data representations
of the same data (e.g. multiple resolutions or fidelities)
that may aid in visual exploration of multi-petabyte imaging data.
It should also be possible to stream the data
progressively to the end-user, displaying an image as soon
as the first data become available. \citet{2014Kitaeff} and
\citet{2014arXiv1401.7433P}  demonstrate the applicability and effectiveness
of such approach on radio astronomy imagery.

\subsection{Capturing indeterminately sized datasets via streaming}
\label{subsection_stream_indeterm}

Frequently there is a need to store data from an instrument or remote
site that is being transmitted over a network. It is common that when
the transfer begins the final size of the dataset is not known. Those
using FITS have handled this by writing such data to a file without
specifying the size of the last dimension in an image or table, and
when the stream is completed, the header is appropriately updated.

Nevertheless, there are applications for which one would like to
access all the other information before the file is complete. This may
be to integrate the data that are being read out, or to monitor
metadata. A library supporting the data format should support such
usage.

\subsection{Virtual and distributed datasets}
\label{subsection_virt_dist_components}

When FITS was created, the `file' (bytes stored on durable physical
medium such as spinning disk or magnetic tape) was more or less the
only way to store and transfer data. The networked solutions which we
enjoy today were absent from the world of astronomy and storage of
astronomical data in databases was unusual. Code was run locally by
experts and the results, if shared at all, were usually only reported
in published papers. In the intervening years, computer and
information technologies have evolved and broadened; we now enjoy many
new means of accessing, providing and storing data. FITS should join
this revolution.

We should start to consider thinking of FITS as a `container' of
astronomical information which is not necessarily a file. Is there
any reason to prevent our FITS `file' from overlaying a portion of a
database? Why not allow FITS to be a wrapper about bytes held within a
distributed mass store such as
iRODS\footnote{\url{http://irods.org}} \citep[see e.g.,][]{2007AGUFMIN13B1214R}
or a cloud?  Similarly, we would
want FITS to contain, and adequately access, data generated by a
service (simulation data, for example). More speculatively, FITS could
itself execute simple stored algorithms to generate a portion of its
data.\footnote{This probably implies the need for a ``FITS language''
to generate these data.}

These use cases are only examples of where we might go and some may be
arguably of limited value. Nevertheless, the generalized use case that
may be derived from all of these is certainly of importance:
a science data storage format should be able to support both local
and remote data access, providing immediate and secure access to data
contained within itself, and providing transparent access to data held
in non-local entities such as cloud storage, databases, services, and
other files.

\section{Discussion -- Lessons Learned}
\label{sec:discussion}

What then are the lessons we might draw from the above analysis
of FITS? Are there any deeper issues and commonalities which thread
throughout these issues? In fact, there are several.

\subsection{Lesson 1. The format should be versioned}

Contrary to the conclusion drawn in \citet{1997ASPC..125..257W}
the first lesson that we may draw is that the format needs to be
versioned.
As we have argued in this paper, we disagree with the premise
that FITS has never undergone significant change and hence, there
is only one version. Without versioning, it becomes a significantly
harder task to write parsers for the format, requiring the software
developer to encompass as many design rules as possible in order
to robustly handle format instances. As the format evolves and
adds new design rules, it only becomes more difficult to write the
next parser and, just as bad, older parsers of the format may fail quietly.
Ultimately, this experience is contrary to the espoused goal of
archivability as one can ultimately never know for certain which
permutation of the format the FITS file being read conforms to.

In contrast, when versioning is present, implementers of parsers
are able to target a subset of the format design, and declare
that within the software so that, should it inadvertently be used
on a version it does not understand, it may fail gracefully and
in a planned manner.

Because it is so important to understanding the design of
the format, versioning metadata should be part of the standard.
The choice to implement this as an optional add-on data model
(such as a FITS convention) is to be avoided. This is because,
without the enforcement of being part of the standard, versioning
is unlikely to be implemented where it is needed most, in the
generation of new instances.

\subsection{Lesson 2. The format should be self-describing}
\label{section_lesson_2}

The next lesson that we may draw is that the format needs to be
``self-describing'' in a machine-readable manner.
We consider a self-describing format to be one where the formatted
instance is capable of conveying and validating the semantic information
it holds where the formatted ``instance'' may be a file, or a
collection of related files or perhaps something more exotic
(see above section~\ref{subsection_virt_dist_components}).

As we have already seen, FITS lacks semantic validation and its
syntactic validation is very limited, achieved only by the creation
of hard-coded rules in software utilities such as \textsc{fitsverify}
(part of the \textsc{ftools} package; \ascl{9912.002})
Furthermore, the limitation on
keyword length to 8 characters all but guarantees that semantic
information within the header is obfuscated. As we have shown,
this in part contributes to the problem of being able to detect,
and implement, multiple data models within a single FITS file
and can lead to the inadvertent creation of informal (and
undetectable) variants. Furthermore, without this validation,
archiving and interchange of information in the format suffers.
It is harder to build robust software systems as any components
involved in the interchange of FITS are unable to adequately
detect, and handle, invalid files fed to it.

The declaration of validation rules should be flexible. In FITS,
where syntactic rules are hard-coded, it is not possible to
declare syntactic rules which check the range or data type of
metadata fields\footnote{Beyond the few canonical keywords
which are part of the FITS standard such as \texttt{GCOUNT}
or \texttt{PCOUNT}} without re-coding the utility.
Ideally, the data format should not rely on hard-coding these rules
in software. Rather, a means to capture and associate the data
model/namespace information with the contents of the formatted
instance, in a machine-readable manner, should be found so that
validation can be possible without human inspection or specifically
written software programs (similar, perhaps, in the way that JSON
or XML formats have schemata).

This approach has additional benefits downstream. First, it will help
to avoid misinterpretation because it is better for the creator
of the data and/or data model to provide the machine-readable information rather
than a downstream programmer. Second, there is a saving in effort
in that the model is done once and need not be repeated by numerous
downstream programmers. Finally, good validation tools will allow the community
to better detect informal variant models and reject them, promoting good
practice.

\subsection{Lesson 3. The format should not limit expression of desired data models}

FITS was originally designed around a data model which contained a single, generic,
two dimensional image and an associated header for metadata. This basic data model
has been expanded to allow more instances (extensions), as well as types of data
(data cubes and tables).
As we discussed in prior sections, working with the basic data model of FITS,
authors have implemented their own data models with ever greater demand being
placed on the type of data (models) FITS may hold.
Holding the line on format changes via the ``once FITS, forever FITS'' doctrine
has been harmful.  Original format design decisions have largely been held onto,
and have limited the expression of new user data models.

There are two general classes of problem which have held back realizing
many needed models. One class concerns the limits created by the format
of the serialization itself. Specifically, we mean the limits on metadata
representation enumerated in section~\ref{subsection_information_representation}
and character encoding in section~\ref{section_char_encoding}.
This class causes difficulties in realizing the WCS model, for example.

The other class of problem is that some needed machinery for data modeling
within the FITS standard itself is missing. Beyond the aforementioned need to
declare models in a machine-readable manner (detailed in
\hyperref[section_lesson_2]{Lesson~2}), this class encompasses a broader range
of issues which include the missing ability to declare byte order,
no standard means to make associations between data and metadata, and
an inability to create data models which extend both
\textit{metadata and the data}. We have already discussed the former two
issues in sections~\ref{section_byte_order} and
\ref{section_data_associations} respectively.  The last issue is related to the fact
that there is no provision in the standard for extending the existing capture
of data itself. For example, if one creates a convention for a new image type
which supports multiple representations of the data (section~\ref{subsection_stream_image})
it is no longer readable by any FITS parser. Other limitations which arise and/or are
unsolvable as a result of this class of problem include no support for
optimization of parallel IO (section~\ref{subsection_parallel_io}), streaming issues
(sections~\ref{subsection_stream_image} and \ref{subsection_stream_indeterm}),
lack of distributed/virtual data representation
and representation of information held in ``non-file'' instances such as in a
database (section~\ref{subsection_virt_dist_components}).

It is critical that this data modeling machinery be integral to the format.
As we have already noted, in many of the above cases it is possible to
create a solution using one or more conventions, but doing so will
result in files which are unparsable by downstream readers which
do not implement these conventions. In addition, there is no apparent
solution, using a FITS convention, than can solve the limitation on the
restricted expression of keyword names, nor can one utilize conventions
to describe how to serialize \textit{data}.

The data format should do as little to impede the expression of data
models.  In practice this means having very few hard-coded rules
within the format itself such as the 2880 byte record or 80 byte card.
Schemata should be capable of describing the layout of \textit{data and metadata}
 and the format serialization should be flexible enough to handle
desired data models and associations between them.

\subsection{Lesson 4. Conventions are not standards}

As currently envisaged, \emph{conventions} have no path of migration
to become part of the FITS standard. There are many useful conventions
that provide features that many people use but no-one can guarantee
that a particular convention will be supported by a FITS reader. A
FITS library cannot simply state that a particular version of the
standard is supported but must also state all the conventions that are
supported. Multiple conventions exist for
continuing long header lines (\emph{multi-spec} and \texttt{CONTINUE})
and for supporting hierarchical headers (\texttt{HIERARCH} and
\emph{record-valued}) but the standard does not have anything to say
as to which convention is preferred. Tile compression is rightfully
thought of as a success for FITS but again tile compression
\citep[e.g.,][]{2007ASPC..376..483S,2009PASP..121..414P} is a
convention and not a standard with no guarantee that a particular
reader will be able to understand the compression scheme.

\textsc{cfitsio}, because it is so heavily relied upon by the
community to implement FITS reader software, may be considered to
be the \emph{de facto} implementation of FITS. This fact, along
with the lack of a migration path for conventions, effectively means
it is also acting as a \emph{de facto} standards body. \textsc{cfitsio}
supports some conventions but not others.
Effectively, the conventions it supports have become mainstream
while others have not. This is the wrong process for
making worthy conventions widespread. We feel that it is important
that the lessons learned from implementing these conventions provide
feedback to the standards process to allow the standard to continue
to grow and evolve over time.

\section{Summary}

The limitations which we have described in this paper are significant
and we have tried to provide an analysis of their deeper origin.
From our investigation, it is clear that FITS suffers from a lack of
sufficient evolution. Original design decisions, such as the header
byte layout and fixed character encoding made a certain sense at the
time FITS was founded. The later enshrinement of the FITS ``Once FITS,
always FITS'' doctrine, which has been utilized to effectively freeze
the format, was a mistake in our opinion. Adherence to the doctrine,
and lack of any means to version the format in a machine-readable manner,
has stifled necessary change of FITS.

More positively, the limitations identified in FITS provide an opportunity
to draw a number of important lessons to be learned from the FITS experience.
Furthermore, we can use our analysis to identify root causes and turn
these into requirements which might be used as goals for future work.
For example, some possible requirements might include:

\begin{itemize}

\item The data format shall be versioned.

\item The data format shall allow for syntactic and semantic validation.

\item The data format standard contain provision for declaration of common
advanced data structures. These structures include non-scalar values, sets,
objects and associations between other metadata and data.

\item The data format shall allow declaration of the character set
(in both metadata and data).

\item The data format shall allow user models to declare new data structures
in data\footnote{FITS conventions are unable to specify new
data structures to capture ``data''. All data models must use existing
n-dimensional array and (binary or ASCII) table data structures}.

\end{itemize}

We should not neglect FITS strengths either. In particular, FITS inclusion of
semantic content (data models) as part of the standard should continue to be
pushed. There are many critical missing data models
(section~\ref{section_crit_data_models}) that, if included in the standard,
would provide compelling reasons to use a data format.

To be clear, we do not wish to recommend specific corrective action
to any particular problem or derived requirement. Instead we hope that
action will flow from constructive community discussion including the analysis
of other astronomical data formats and any lessons learned from their use
and construction.
We may anticipate that the form of the possible resolutions to problems in FITS may
involve moving existing FITS conventions into the core standard, modification of
the FITS standard to remove limitations, or even transferring the FITS data model over
into a new serialization.

Our effort will also continue. We plan to extend the work started
here in a future paper in which we will gather use cases or ``lessons learned''
which also show FITS strengths, and gleaning the same from other data formats.
From these we plan to extract and publish an overview of the requirements for
a modern astronomical data format.


\section{Acknowledgments}

The authors wish to thank Bill Pence and Rob Seaman for many insightful
discussions.  We thank Bob Hanisch and Michael Wise for providing
helpful comments that improved the contents of this paper.

This research has made use of NASA's Astrophysics Data System.

\bibliographystyle{model2-names}
\bibliography{acfits}








\end{document}